\documentstyle[twocolumn,aps,prl,overcite]{revtex}

\input epsf

\tighten

\begin{document}
\title{A Dynamical Solution to the Problem of a Small Cosmological Constant
and Late-time Cosmic Acceleration}

\author{C. Armendariz-Picon,$^1$ V. Mukhanov$^1$ and Paul J. Steinhardt$^{2}$}

\address{
$^1$Ludwig Maximilians Universit\"{a}t, Sektion Physik, M\"{u}nchen, Germany \\
$^2$Department of Physics, Princeton University, Princeton, NJ 08540}

\maketitle

\begin{abstract}
Increasing evidence suggests that most of the energy density of the universe
consists of a dark energy component with negative pressure, a ``cosmological
constant" that causes the cosmic expansion to accelerate. In this paper,
we address the puzzle of why this component comes to dominate the universe
only recently rather than at some much earlier epoch. We present a class of
theories based on an evolving scalar field
where the explanation is based  entirely on internal dynamical properties of
the solutions. In the theories we consider,  the dynamics causes the
scalar field to   lock automatically into a negative pressure state at the onset of
matter-domination such that the present epoch is the earliest possible time, 
consistent with nucleosynthesis
restrictions, when it can start to dominate.  
\end{abstract}


{\it Introduction.} Observations of large scale structure, searches for Type
Ia supernovae, and measurements of the cosmic microwave background
anisotropy all suggest that the universe is undergoing cosmic acceleration
and is dominated by a dark energy component with negative pressure.\cite
{bahcall} The dark energy may consist of a cosmological constant (vacuum
density) or quintessence,\cite{quint} such as a 
scalar field with negative pressure. In either case, a key
challenge is the ``cosmic coincidence'' problem: Why is it that the vacuum
density or scalar field dominates the universe only recently? Until now,
either cosmic initial conditions or model parameters (or both) had to be
tuned to explain the low density of the dark energy component.

In this paper, we explore a new class of scalar field models with novel
dynamical properties that avoid the fine-tuning problem altogether. A
feature of these models is that the negative pressure results from the
non-linear kinetic energy of the scalar field, which we call, for brevity,
$k$-field or  $k$-essence. (This
consideration is inspired by earlier studies of $k$-inflation, kinetic
energy driven inflation.\cite{kinfl,GaMu,Chiba}). As we will show, for a
broad class of theories, there exist attractor solutions which determine
the equation-of-state of $k$-essence during different epochs depending on
the equation-of-state of the background. Effectively, the scalar field changes
its speed of evolution in dynamic response to changes in the background
equation-of-state. During the radiation-dominated
epoch, $k$-essence is led to be subdominant and to mimic the equation-of-state
of radiation. Hence, the ratio of $k$-essence to radiation density remains
fixed. When the universe enters the dust-dominated epoch, though, $k$-essence
is unable to mimic the dust-like equation-of-state for dynamical reasons.
Instead, the energy decreases rapidly by several orders of magnitude and
freezes at a fixed value. After a period (typically corresponding roughly to
the current age of the universe), the field overtakes the matter density and
drives the universe into cosmic acceleration. Ultimately, the $k$-essence
equation-of-state slowly relaxes to an asymptotic value between 0 and -1.
(The reader may wish to sneak a peek at Fig. 3 which illustrates the
behavior in a specific numerical example.)

The scenario bears some likeness to the quintessence ``tracker models''
discussed by Zlatev {\it et al.}\cite{Zlatev,track2}, but is in fact very
different from them. For a certain class of tracker
potentials, the quintessence scalar field converges to an attractor solution
in which the energy density in the quintessence field mimics the
equation-of-state of the background (matter or radiation)
energy density. However, for tracker
potentials, it does not make a difference what the equation-of-state of the
background is. Only after the scalar field passes a certain critical value does the 
quintessence develop  a negative pressure, and, then, the energy density becomes
fixed.
The weakness in this model is that the energy
density for which the pressure becomes negative is set
by an adjustable parameter,  which has to be fine
tuned to explain why cosmic acceleration is happening at the present epoch in the
history of the universe.

The distinctive feature of
 $k$-essence models is that tracking of the background energy 
 density can only occur in the radiation epoch.
 At the matter-radiation equality, a sharp 
 transition of $k$-essence  from positive to
negative pressure is automatically 
triggered by dynamics.
The  $k$-essence
cannot dominate before matter-radiation equality because it is exactly tracking
the radiation background. It also cannot  dominate immediately after
dust-domination because its energy density necessarily drops several orders of
magnitude at the transition to dust-domination. However, since its
energy density decreases more slowly than the matter density as the
universe expands, $k$-essence must
dominate not too long thereafter, at roughly the current epoch.
The resolution of the 
cosmic coincidence problem boils down to 
the fact that we live at the ``right time'' 
after matter-radiation equality.

As noted above,
the remarkable behavior comes at the cost of introducing a non-linear 
kinetic energy density functional of the scalar field and
adjusting it to obtain the desired attractor behavior.
This kind of action may describe a fundamental
scalar field or be a low-energy effective action. For example,
in string and supergravity theories, non-linear kinetic 
terms appear generically in the effective action describing moduli and
massless degrees of freedom (superpartners) due to higher order 
gravitational corrections to the Einstein action.\cite{Gross,Polch}
The attractor behavior of our models
relies on certain broad conditions on the form of these terms. Our initial 
examples are admittedly contrived for the purposes of numerical illustration.
A systematic study of model-building will appear in a forthcoming
paper,\cite{ArMuSt} although, having seen here  the relatively simple
basic principles, the reader should be equipped to explore  more
attractive and better-motivated forms.

{\it Equations. }In the theories we consider \ the Lagrangian density for
$\varphi $ is taken to be 
\begin{equation}\label{lag}
{\cal L=}-\frac{1}{6}R+\frac{1}{\varphi ^{2}}\tilde{p}_{k}(X)+{\cal L}_{m}
\end{equation}
where $R$ is the Ricci scalar, $X\equiv \frac{1}{2}(\nabla \varphi )^{2}$ ,  
${\cal L}_{m}$ is the Lagrangian density for dust and radiation and we use
units where  $8\pi G/3=1$. 
  The
energy density of the  $k$-field $\varphi $ is $\rho _{k}=(2X\tilde{p}_{,X}-
\tilde{p})/\varphi ^{2}$; the pressure is $p_{k}=\tilde{p}/\varphi ^{2}$;
and the speed of sound of $k$-essence is $c_{s}^{2}=p_{k,X}/\rho
_{k\,,X}$, where the subscript  means derivative with respect to $X$.\cite
{kinfl,GaMu} 

The attractor behavior can be explained most easily by changing variables
from $X$ to $y=1/\sqrt{X}$ and rewriting the $k$-field Lagrangian as: 
\begin{equation}
{\cal L}_{k}{\cal =}\tilde{p}_{k}(X)/\varphi ^{2}\equiv g(y)/\varphi ^{2}y.
\end{equation}
In this case, the energy density and pressure are $\rho
_{k}=-g^{\prime }/\varphi ^{2}$ and $p_{k}=g/\varphi ^{2}y$,
where prime indicates derivative with respect to $y$. The
equation-of-state is 
\begin{equation}
w_{k}\equiv p_{k}/\rho _{k}=-g/yg^{\prime }  \label{eos}
\end{equation}
and the sound speed is 
\begin{equation}
c_{s}^{2}=\frac{p_{k}^{\prime }}{\rho _{k}^{\prime }}=
\frac{g-g^{\prime }y}{g^{\prime \prime }y^{2}}.
\end{equation}

In order to have a sensible, stable theory, we require $\rho _{k}>0$
and $c_{s}^{2}>0$. These conditions are satisfied if $g^{\prime }<0$ and $%
g^{\prime \prime }>0$ in the region where $p_{k}^{\prime }$ is positive.
Therefore, a general, convex, decreasing function $g(y)$, such as shown in
Fig. 1, satisfies these necessary conditions. Using the Friedmann equation:
$H^{2}=\rho _{tot}=\rho _{k}+\rho _{m}$ ,where
$\rho _{m}$ is the energy density of  matter (radiation and dust), and the
energy conservation equations, $\dot{\rho}_{i}=-3\rho
_{i}(1+w_{k})$ for the $k$-essence $\left( i\equiv k\right) $ and matter $\left(
i\equiv m\right) $ components, we obtain the following equations of motion 
\begin{eqnarray} \label{master1}
\dot{y} & = & \frac{3}{2}\frac{(w_{k}\left( y\right) -1)}{r^{\prime }\left(
y\right) }\left[ r\left( y\right) -\sqrt{\frac{\rho _{k}}{\rho
_{tot}}}\right] 
\\ \label{master2}
\left( \frac{\rho _{k}}{\rho _{tot}}\right) ^{\displaystyle .}&=&3\frac
{\rho _{k}}{\rho _{tot}}\left( 1-\frac{\rho _{k}}
{\rho _{tot}}\right) \left( w_{m}-w_{k}\left( y\right) \right) ,
\end{eqnarray}
where
\begin{equation}\label{master3}
\,r(y)\equiv \left( -\frac{9}{8}g^{\prime }\right) ^{1/2}y\left(
1+w_{k}\right) =\frac{3}{2\sqrt{2}}\frac{\left( g-g^{\prime }y\right) }
{\sqrt{-g^{\prime }}},
\end{equation}
and dot denotes derivative with respect to $N\equiv {\rm ln}\,a.$
These are the master equations describing the dynamics of $k$-essence
models. Once some general properties of $g(y)$ are specified, the attractor
behavior described in the introduction follows from these coupled equations.

{\it Dynamics. }We are seeking a tracker solution $y(N)$ in which the
$k$-essence equation-of-state is constant and exactly equal to the 
background equation-of-state, 
$w_{k}(y\left( N\right) )=w_{m}$, and the ratio $\rho
_{k}/\rho _{tot}$ is fixed. Generically, this requires $y(N)$ be
a constant $y_{tr}$ and therefore $\rho _{k}/\rho _{tot}=r^{2}(y_{tr})$.
The last condition can only be satisfied if $r(y_{tr})$ is less than unity.
Hence, given a convex function
such as shown in Fig. 1, we can first identify those ranges of $y$ where
$r(y)$ is greater than unity or less than unity. In ranges where $r(y)$ is
less than unity, we can seek values of $y$ where $w_{k}$ in Eq.~(\ref{eos})
is equal to $w_{m}$, the equation-of-state of the matter or radiation. 
The value of $y_{tr}$ changes depending on the epoch and $w_m$.
These are the attractor solutions. In ranges where $r(y)
$ exceeds unity, there are no attractor solutions.

\begin{figure}
 \epsfxsize=3.3 in \epsfbox{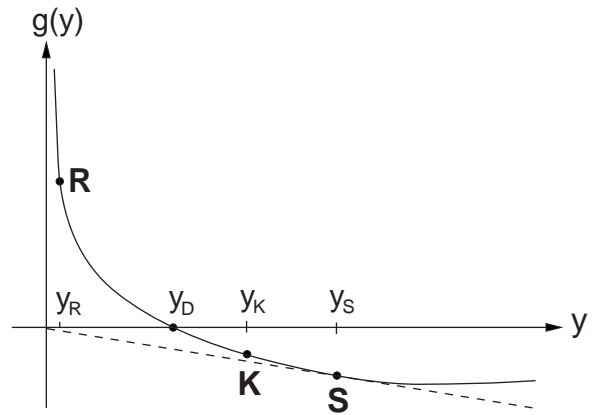}
\caption{ A plot of $g(y)$ vs. $y$ (see Eq. (3) for definition) indicating
the points discussed in the text. {\bf R} corresponds to the attractor
solution during the radiation-dominated epoch; {\bf S} is the de Sitter
attractor at the onset of matter-domination; and {\bf K} is the attractor as
$k$-essence dominates. For our range of $g(y)$, there is no dust-like
attractor solution at $y=y_D$.}
\end{figure}
Although there is no dust attractor, it is quite possible for there to be a
radiation attractor. The radiation attractor corresponds to positive
pressure, so it can be located only at  $y<y_{D}$. Hence, we must have $g(y)$
such that $r(y)$ is less than unity for some range at $y<y_{D}$ which
includes some point $y_{R}$ where $w_{k}(y_{R})=1/3$. During the
radiation-dominated epoch, the ratio of $k$-essence to the total density
remains fixed on this attractor and equal to $\left( \rho
_{k}/\rho _{tot}\right) =r^{2}(y_{R}).$

In Fig. 1, the pressure $p_{k}=g/\varphi ^{2}y$ is positive above the
$y$-axis, and negative below the $y$-axis. The dust equation-of-state $p_{k}=0$
can only be obtained at $y=y_{D}$ where $g(y)$ goes through zero. However,
this point can be an attractor, only if, the second condition, $r(y_{D})<1,$
is satisfied. If it so happens that $r(y_{D})>1$, then there is no dust
attractor in the matter-dominated epoch. This is precisely what we want
for our scenario, and this is possible for a broad class of functions $g$.

If $g$ possesses a radiation attractor but no dust attractor, what happens
at dust-radiation equality? To answer this question let us study the
solutions of the master equations Eqs.~(\ref{master1}-\ref{master3}) in two
limiting cases, when the energy density of $k$-essence is either much
smaller or much greater than the matter energy density. If $\rho
_{k}/\rho _{m}\ll 1,$ one can neglect the last term in the equation
(\ref{master1}) and it is obvious that $y\left( N\right) \simeq y_{S},$ where 
$y_{S}$ satisfies the equation $r\left( y_{S}\right) =  0$, is an
approximate solution of the equations of motion. The point $S$ satisfies
$g(y_{S})=g^{\prime }(y_{S})y_{S}$, so the tangent of $g$ at $y_{S}$ passes through
the origin, as shown in Fig. 1. Since $r\propto (1+w_{k}),$ the equation of
state of $k$-essence at $y_{S}$ corresponds to $w_{k}\left( y_{S}\right)
\approx -1;$ we call this solution the de Sitter attractor and denote it by 
{\bf S} in Fig. 1. From Fig. 1, it is clear that $y_{S}$ nearly always
exists for convex decreasing functions $g$. We stress that $y\left( N\right)
\approx y_{S}$ is an approximate solution to the equations-of-motion only
when matter strongly dominates over $k$-essence. So, if $\rho _{k}$
during the radiation dominated epoch is significantly less than the
radiation density which it tracks, which is both typical and required to
satisfy nucleosynthesis constraints, then $k$-essence proceeds to the de
Sitter attractor immediately after dust-radiation equality.\cite{ArMuSt}

As the transition to dust-domination occurs, $\rho _{k}$ first drops
to a small, fixed value, as can be simply understood. Suppose that $\left(
\rho _{k}/\rho _{tot}\right) _{R}=r^{2}\left( y_{R}\right)
=\alpha <10^{-2}$ during the radiation dominated epoch, where the bound is
set by nucleosynthesis constraints. From the equation of state,
Eq.~(\ref{eos}), we have the relation: $g(y_{R})=-g^{\prime
}(y_{R})y_{R}/3$. The condition $r(y_{D})\geq 1$ is required in order to
have no dust attractor solution. Combining these relations, we obtain: 
\begin{equation}
\frac{g_{R}^{\prime }y_{R}^{2}}{g_{D}^{\prime }y_{D}^{2}}\leq
\frac{9}{16}\alpha <10^{-2}  \label{r1}
\end{equation}
On the other hand, it is apparent from Fig. 1 that $-{g^{\prime }}_{R}>-
{g^{\prime }}_{D}$, so $y_{R}\ll y_{D}$ if $\alpha \ll 1$. In particular, the
tangent at $y_{D}$ falls below $g(y_{R})$, so $g_{D}^{\prime
}(y_{R}-y_{D})\approx -y_{D}g_{D}^{\prime }\leq g(y_{R})=-y_{R}g_{R}^{\prime
}/3$. Using this relation, we obtain 
\begin{equation}
\frac{y_{R}}{y_{D}}\leq \frac{3}{16}\alpha <2\cdot 10^{-3}\text{ \ and \ \ \
\ }\frac{g_{D}^{\prime }}{g_{R}^{\prime }}\leq \frac{\alpha }{16}<7\cdot
10^{-4}.  \label{r2}
\end{equation}
Since $\rho _{k}=-g^{\prime }/\varphi ^{2}$ and $\left| g^{\prime
}\left( y_{S}\right) \right| \leq \left| g^{\prime }\left( y_{D}\right)
\right| $ we conclude that after radiation domination, when the $k$-field
reaches the vicinity of the {\bf S}-attractor, the ratio of energy densities
in $k$-essence and dust does not exceed $(\rho _{k}/\rho
_{tot})_{R}\times g_{D}^{\prime }/g_{R}^{\prime }$;  that is, $\rho
_{k}/\rho _{dust}<\alpha ^{2}/16<7\cdot 10^{-6}.$ Hence, provided
$(\rho _{k}/\rho _{tot})_{R}\leq 10^{-2}$ at dust-radiation
equality, the $k$-essence field loses energy density on its way to the
{\bf S}-attractor down to a value below $7\times 10^{-6}$.

By definition, the {\bf S}-attractor is one in which $w\approx -1$ and the
energy density is nearly constant. Hence, once $\rho _{k}$ has
reached its small but non-zero value, it freezes. In the further evolution
of the universe, the matter density decreases, but the $k$-essence energy
density remains constant, eventually overtaking the matter density of the
universe. Note that, as $\rho _{k}$ approaches $\rho _{m}$,
the condition $\rho _{k}/\rho _{m}\ll 1$ is necessarily
violated and a new attractor solution is found for the case where $k$
-essence itself dominates the background energy density. This attractor is
denoted {\bf K} in Fig. 1.

To prove that that the {\bf K}-attractor exists, we consider the master
equations, Eqs.~(\ref{master1}-\ref{master3}), in the limit where
$\rho _{k}/\rho _{tot}\rightarrow 1$. If $y_{K}$ satisfies
the equation $r\left( y_{K}\right)=1,$ then $y\left( N\right) \simeq y_{K}$ 
is an approximate solution of the equations of motion. When dust is
not a tracker, there always exists  a  unique
attractor $y_{K}$  in the interval $y_{D}<y<y_{S}$.\cite{ArMuSt}
To prove this, note that, within this
interval, the function $r(y)$ has a negative derivative. Recall that $r(y_S)
=0$ (definition of {\bf S}-attractor) and $r\left( y_{D}\right) >1$ (to
avoid a dust attractor). Since $r\left( y\right) $ is a monotonically
decreasing, continuous function, there exists a unique point $y_{K}$
($y_{D}<y_{K}<y_{S})$)  where $r(y)$
becomes equal to unity. At $y>y_{D}$ the
pressure of $k$-essence is negative. Hence, generically the {\bf K }
attractor, located near $y_{K},$ describes a universe dominated by a
negative pressure component which induces power-law cosmic acceleration. As
acceleration proceeds, $\rho_k$ increasingly dominates and $y
\rightarrow y_K$.

Following along using Fig. 1, the dynamics can be summarized as follows:  
$k$-Essence is attracted to $y=y_R$ during the radiation dominated epoch;
at matter-domination, the energy density drops sharply as
$k$-essence skips past $y=y_D$, because there is no dust attractor, and heads
towards $y \approx y_S$. The energy density $\rho_k$ freezes and,
after a period, overtakes the matter density. As it does so, $y$ relaxes
towards $y_K$. In this scenario, our current universe would be making the
transition from $y_S$ to $y_K$. All this occurs for generic $g(y)$
satisfying broad conditions on its first and second derivatives. If the
ratio of $\rho_k$ to the radiation density is near the maximum
allowed by nucleosynthesis (roughly equipartition initial conditions), the
scenario predicts that the $\rho_k$ dominates by the present epoch.


{\it Numerical results.} We have verified these analytic predictions
 numerically for a wide class of $g(y)$.  As as a strategy, we
 look for forms which are roughly 
 linear, 
 \begin{equation}
 g\left( y\right) \approx - \frac{1}{3}g_R^{\prime} y_R +
g_{R}^{\prime }(y- y_{R})+
 O\left( \left( y-y_{R}\right) ^{2}\right)   \label{radass}
 \end{equation}
in the vicinity of radiation attractor {\bf R} and parabolic 
\begin{equation}
g\left( y\right) \approx  \frac{g_{D}^{\prime }y_{D}}{y_{D}^{2}-y_{S}^{2}}
\left( y-y_{D}\right) \left( y-\frac{y_{S}^{2}}{y_{D}}\right) +...
\label{dustass}
\end{equation}
in the region $y_{D}\leq y\leq y_{S}$. 
  One can easily check, that the points 
$y_{R}$, $y_D$ and $y_{S}$ here are, by construction, the places where the
corresponding attractors are located and $g_{R}^{\prime },g_{D}^{\prime }$
are the derivatives of $g$ at the appropriate points. The results are not
sensitive to the precise form of $g$ that interpolates between these
regimes. The main constraints are that the attractor 
solution have  a small ratio of $\rho_k/\rho_{tot}$ 
during the radiation epoch and that there is no dust attractor.

\begin{figure}
\epsfxsize=3.3 in \epsfbox{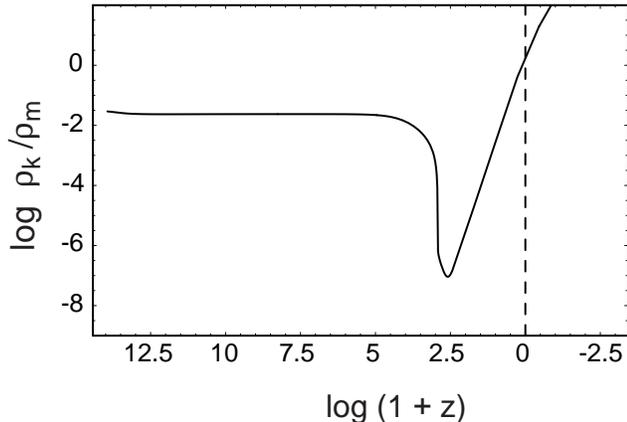}
\caption{ The ratio of $k$-essence energy density,  $\rho_k$, to the  density 
in radiation and matter, $\rho_m$, vs.
red shift. At the present epoch (dashed line), $\Omega_k \approx 0.7$. }
\end{figure}

\begin{figure}
\epsfxsize=3.3 in \epsfbox{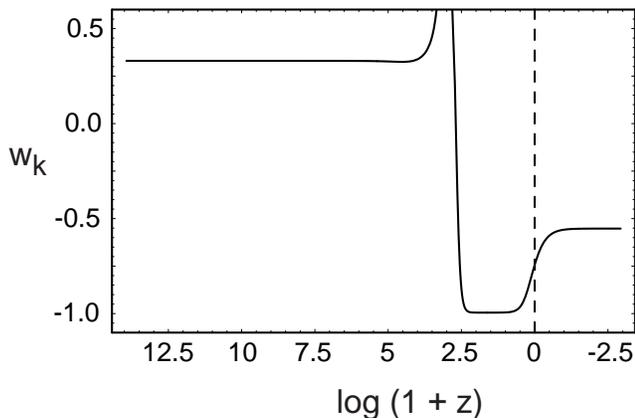}
\caption{ The $k$-essence equation-of-state vs. red shift. The three
attractors in the radiation-, matter-, and $k$-essence-dominated
epochs are evident. At the present epoch, $w_k  \approx -0.77$. }
\end{figure}

For illustrative purposes, we have used these
principles to  obtain a sample  $g(y)$ and  have
transformed it to a form for $\tilde{p}(X)$ in the action,
 Eq.~(\ref{lag}):
\begin{equation}
  \tilde{p}_k(X)=-2.01 + 2 \sqrt{1 + X} + 0.03 ( a X)^3 - (b X)^4,
\end{equation}
where $a= 10^{-5}$ and $b= 10^{-6}$.  For small values of $X$, after a field
redefinition, this Lagrangian density reduces to  one equivalent 
to  a canonical scalar
field with exponential potential (curiously, a tracker model\cite{Zlatev}).
 The distinctive  dynamical attractor in our models 
 relies deviations from linearity at large $X$. 

The results of a numerical integration are presented in Figs. 2 and 3. 
We see that  $k$-essence tracks  the radiation ($w_k \approx 1/3$) during the 
radiation-dominated epoch.  
Then, at the onset of matter-domination,
$w_k $ starts to change and the energy density of $k$-essence suddenly drops
by several orders of magnitude becoming of the order of 10$^{-7}$ times the
critical density at red shifts about $z$ $\simeq 1000$ as the
{\bf S}-attractor is approached and $w \rightarrow -1$. 
At about red shift $z\sim
3-5,$ $\rho_k$ becomes non-negligible and $w_{k}$ starts to increase,
ultimately reaching $w_{k}\simeq -0.77$ at $z=0$. 
The ratio
of the $k$-essence energy density to the critical density today is $\Omega_k
\approx 0.74$. In the future, $w_{k}$ in this model has to approach the value 
$-0.55$, corresponding to the {\bf K}-attractor solution, and the universe
will enter the period of power law $k$-inflation.

{\it Summary.} In this paper, we have presented a scenario in which cosmic
acceleration occurs late in the history of the universe due to an inevitable
sequence of events caused by attractor dynamics. We view the present work as
a demonstration of principle; hence, we have emphasized general conditions
and an analytic understanding of the scenario. The specific example 
illustrated in this paper is admittedly  complex, composed to illustrate the
concept, but we know of no fine-tuning or other requirement that poses a
barrier to finding simpler and better-motivated forms. By changes of
variables and other simple techniques, one can quickly enlarge the class of 
actions considered here, which may suggest other types of attractive
models.

A prediction of $k$-essence models that distinguishes them from models based on
tracker potentials\cite{track2} is that $w_{k}$ is in the process
of increasing today from -1 towards its asymptotic value at the {\bf K}
attractor, whereas, for trackers, $w_{k}$ is undergoing a transition from
$w\approx 0$ towards $w=-1$. A consequence is that the effective value of
$w_{k}$ for $k$-essence  -- 
that is, the $\Omega _{k}$ weighted average of $w_{k}$
between the present and $z=1$ --  can be significantly lower than for the
tracker potential case, which is bounded below by $%
w_{eff}\approx -0.75$.\cite{track2} 
In the numerical example above, the effective
$w_{eff}=-0.84$, for example. 
The current supernovae data suggest a lower value of $%
w_{k}$ more consistent with $k$-essence.\cite{sndata} Of course, the
$k$-essence range for $w_{k}$ is more difficult to distinguish from a
cosmological constant ($w=-1$).

In future work,\cite{ArMuSt} 
we discuss model-building using further examples and
generalizations. We also explore interesting
variations of the dynamical  scenario with different kinds of 
attractors, including some
which can lead to different long-term future outcomes, such as a return to a
pressureless, unaccelerated expansion in the long-term future.

This work was initiated
at the Isaac Newton Institute for Mathematical Sciences;
we thank the organizers and staff of the Institute for
their support and kind hospitality.
This work was supported in part  by the ``Sonderforschungsbereich 
375-95 f\"ur Astro-Teilchenphysik" der Deutschen
Forschungsgemeinschaft (C.A.P. \& V.M.) and by Department of Energy grant
DE-FG02-91ER40671 (Princeton) (P.J.S.).





\begin{references}
\bibitem{bahcall}  See, for example, N. Bahcall, J.P. Ostriker. S.
Perlmutter, and P.J. Steinhardt, {\it Science} {\bf 284}, 1481-1488, (1999)
and references therein.

\bibitem{quint}  R.R. Caldwell, R. Dave and P.J. Steinhardt, {\it Phys. Rev.
Lett.} {\bf 80}, 1582 (1998); P. G. Ferreira and M. Joyce, {\it Phys. Rev.
Lett.} {\bf 79}, 4740 (1997); J. Frieman, C. Hill, A. Stebbins, I. Waga, 
{\it Phys. Rev. Lett.} {\bf 75} 2077 (1995); P.J.E. Peebles and B. Ratra, 
{\it Ap. J. Lett.} {\bf 325}, L17 (1988); B. Ratra and P.J.E. Peebles,
{\it Phys. Rev. D} {\bf 37}, 3406 (1988).

\bibitem{kinfl}  C. Armendariz-Picon, T. Damour, and V. Mukhanov, {\it Phys.
Lett. B}{\bf 458}, 209 (1999).

\bibitem{GaMu} J.Garriga, V.Mukhanov, {\it Phys. Lett. B}{\bf 458}, 219 (1999).

\bibitem{Chiba}  T.Chiba, T.Okabe and M.Yamaguchi, astro-ph/9912463.

\bibitem{Zlatev}  I. Zlatev and P.J. Steinhardt, {\it Phys. Lett. B}{\bf 459},
570-574 (1999).

\bibitem{track2}  I. Zlatev, L. Wang and P.J. Steinhardt,
{\it Phys. Rev. D}{\bf 59}, 123504 (1999).


\bibitem{Gross} D. Gross and E. Witten,  
{\it Nucl. Phys. B}{\bf 277},
1 (1986).

\bibitem{Polch} J. Polchinski, {\bf Superstrings, Vol II}, 
(Cambridge U. Press, Cambridge, 1998), Ch. 12.


\bibitem{ArMuSt}  C. Armendariz-Picon, V. Mukhanov, and P.J. Steinhardt, in
preparation.



\bibitem{sndata}  P.M. Garnavich, {\it et al.}, {\it Astrophys. J.} {\bf 509},
74 (1998); S. Perlmutter, M.S. Turner, and M. White, {\it Phys. Rev. Lett.}
{\bf 83} 670-673 (1999);
\end{references}
\end{document}